\documentclass[aps,prd,twocolumn,superscriptaddress]{revtex4}
\usepackage{epsfig,epsf}
\usepackage{amsmath}
\usepackage{amsthm}
\usepackage{amsfonts}
\usepackage{amssymb}
\usepackage{dsfont}
\usepackage{multirow}
\usepackage{appendix}
\usepackage{slashed}
\usepackage[active]{srcltx}
\usepackage{psfrag}
\usepackage{subfigure}

\begin{document}

\title{Thermal properties of the exotic $X(3872)$ state via QCD sum rule}
\date{\today}
\author{E.~Veli~Veliev}
\affiliation{Department of Physics, Kocaeli University, 41380 Izmit, Turkey}
\affiliation{Education Faculty, Kocaeli University, 41380 Izmit, Turkey}
\author{S.~G\"{u}nayd{\i}n}
\affiliation{Department of Physics, Kocaeli University, 41380 Izmit, Turkey}
\author{H.~Sundu}
\affiliation{Department of Physics, Kocaeli University, 41380 Izmit, Turkey}

\begin{abstract}
In this work we investigate the $X(3872)$ meson with quantum numbers $%
J^{PC}=1^{++}$ in the framework of the thermal QCD sum rules
method. We use a diquark-antidiquark current with the
corresponding quantum numbers and calculate the two-point
correlation function including contributions of non-perturbative
condensates up to six dimensions. Analysis of the obtained thermal
sum rule allows us  to study contributions of a medium to the mass
and coupling constant of the $X(3872)$ resonance. Our numerical
calculations demonstrate that the mass and the meson-current
coupling constant are insensitive to the variation of  temperature
up to $T=110~MeV$ , however after this point;  they start to fall
by increasing the temperature. At deconfinement temperature, the
meson-current coupling constants attain roughly to $34\%$ of the
vacuum value.
\end{abstract}

\maketitle

\section{Introduction}

Understanding the non-perturbative properties of QCD is one of the
most difficult and intriguing research topics of the in strong
interactions. The investigation of the hadron spectrum can play an
important role in achieving this goal. According to QCD, not only
traditional mesons and baryons, but also exotic particles such as
glueball, hybrid and multiquark states can be observed. In
the last decade, observation of charmonium-like \cite%
{Choi:2003ue,Acosta:2003zx,Abazov:2004kp}, bottomonium-like \cite%
{Adachi:2011ji,Abe:2007tk} and the pentaquark states
\cite{Aaij:2015tga} in experiments and detailed examination of
these states have revealed significant information on exotic
particles. The observations of these unconventional states with
some properties beyond the standard quark model have motivated
different theoretical interpretations such as molecule and
tetraquark models \cite%
{Swanson:2006st,Richard,Zhu:2007wz,Godfrey:2008nc,Nielsen:2009uh,Chen:2010ze, Brambilla:2010cs,Albuquerque:2013ija,Liu:2013waa,Bodwin:2013nua,Nielsen,Braaten, Esposito:2014rxa,Chen,Briceno,Wang:2016mmg,Ali:2016gli,Wang:2016gxp,Wang:2016tzr}%
. The examination of exotic particles is very important to understand the
investigation of heavy ion collisions. But for understanding the
investigation of heavy ion collisions, we need to know the thermal
properties of these particles. For this reason, it is important to examine
the exotic particles in the medium.

In 2003 the Belle collaboration announced the discovery of the X(3872)
particle \cite{Choi:2003ue}. This particle was confirmed shortly thereafter
by the CDF \cite{Acosta:2003zx}, D0 \cite{Abazov:2004kp}, BaBar \cite%
{Aubert:2004ns,Aubert:2005rm} and LHCb \cite{Aaij:2014ala} collaborations
 by
analyzing the $B^-\rightarrow \pi^- [X\rightarrow J/\psi \pi^+\pi^-]$, $%
B^-\rightarrow K^- [X\rightarrow J/\psi \pi^+\pi^-]$, $B^+\rightarrow K^+
[X\rightarrow \psi(2S)\gamma]$ and $B^+\rightarrow K^+ [X\rightarrow \psi
\gamma]$ decays.

In the last decades, QCD sum rules have successfully been used to
investigate different properties of the conventional mesons and
baryons as reviewed in Refs.
\cite{Shifman,Reinders,Colangelo:2000dp,Narison:2002pw}. The
thermal version of QCD sum rules has been successfully used to
study the thermal properties of mesons
\cite{Bochkarev,Hatsuda,Morita,Gubler,Azizi:2015oxa,
Azizi:2016ddw,Azizi:2015ona,Azizi:2015qva,Azizi:2010zza,Veliev:2010gb,
Azizi:2014maa,Veliev:2014tca,Yazici:2015tqa,Veliev:2008zi} as a
reliable and well-established approach. The QCD sum-rules have
been also extended to
investigate exotic hadrons \cite%
{Agaev:2017foq,Agaev:2017uky,Agaev:2016dsg,Agaev:2016ifn,Agaev:2016srl,Sundu:2016oda,Abreu:2017nuc,Cho:2017dcy, Abreu:2016kbd,Abreu:2016qci,Wang:2017jtz,Wang:2016dcb,Wang:2016wwe}%
. In this article we use the Thermal QCD sum rules method for exploration of
the $X(3872)$ resonance with quantum numbers $J^{PC} = 1^{++}$. We consider
it as a diquark-antidiquark bound state. By using relevant interpolating
current we calculate the two-point correlation function including
contributions of nonperturbative condensates up to six dimensions. Equating
the expression of the correlation function obtained using the operator
product expansion (OPE) and its hadronic representation we derive thermal
QCD sum rules for parameters of the $X(3872)$ state.

This work is organized in the following manner. In Sec.\
\ref{sec:Mass} we derive the thermal sum rules to calculate mass
and coupling constant of the resonance $X(3872)$. Section
\ref{sec:Num} is devoted to numerical analysis, where we write
down values of parameters used in computations, and also present
our results for the mass and coupling constant of the $X(3872)$
state. The appendix contains the explicit expression of the
two-point thermal spectral density $\rho^{\mathrm{QCD}}(s, T)$.


\section{Mass and coupling constant of the $X(3872)$ state at finite
temperature}

\label{sec:Mass}

To calculate the mass and coupling constant of the $X(3872)$ state in the
framework of the thermal QCD sum rules we start from the correlation
function
\begin{equation}
\Pi _{\mu \nu }(q,T)=i\int d^{4}xe^{iq\cdot x}\langle \mathcal{T}\{J_{\mu
}(x)J_{\nu }^{\dag }(0)\}\rangle ,  \label{eq:CorrF1}
\end{equation}%
where $J_{\mu }(x)$ is the interpolating current of the $X(3872)$ state, $T$
is the temperature and $\mathcal{T}$ indicates the time ordering operator.
The thermal average of any operator $A$ in thermal equilibrium can be
expressed as:
\begin{equation}  \label{eqn2}
\langle A\rangle=Tr e^{-\beta H}A/Tr e^{-\beta H}, \\
\end{equation}
where $H$ is the QCD Hamiltonian, and $\beta=1/T$ is the inverse of the
temperature $T$.

We consider the $X(3872)$ state as the resonance with quantum numbers $%
J^{PC} = 1^{++}$. Then in the diquark-antidiquark model the current $J_{\mu}
(x)$ is expressed by the following expression \cite{Matheus:2006xi}
\begin{eqnarray}
&&J_{\mu }(x)=\frac{i\epsilon \tilde{\epsilon}}{\sqrt{2}}\left\{ \left[
q_{a}^{T}(x)C\gamma _{5}c_{b}(x)\right] \left[ \overline{q}_{d}(x)\gamma
_{\mu }C\overline{c}_{e}^{T}(x)\right] \right.  \notag \\
&&\left. +\left[ q_{a}^{T}(x)C\gamma _{\mu }c_{b}(x)\right] \left[ \overline{%
q}_{d}(x)\gamma _{5}C\overline{c}_{e}^{T}(x)\right] \right\} ,
\label{eq:CDiq}
\end{eqnarray}%
where $q$ is one of the light $u$ or $d$ quarks. Here we have introduced the
short-hand notations $\epsilon =\epsilon _{abc}$ and $\tilde{\epsilon}%
=\epsilon _{dec}$. In Eq.\ (\ref{eq:CDiq}) $a,b,c,d,e$ are color indexes and
$C$ is the charge conjugation matrix.

In order to derive QCD sum rule expression we first calculate the
correlation function in terms of the physical degrees of freedom. Performing
integral over $x$ in Eq.\ (\ref{eq:CorrF1}), we get
\begin{equation*}
\Pi _{\mu \nu }^{\mathrm{Phys}}(q,T)=\frac{\langle 0|J_{\mu }|X(q)\rangle_T
\langle X(q)|J_{\nu }^{\dagger }|0\rangle_T }{m_{X}^{2}(T)-q^{2}}+ \ldots,
\end{equation*}
where $m_{X}(T)$ is the temperature-dependent mass of $X(3872)$, and dots
stand for contributions of the higher resonances and continuum states. We
define the temperature-dependent meson-current coupling constant $f_{X}(T)$
through the matrix element
\begin{equation}
\langle 0|J_{\mu }|X(q)\rangle_T =f_{X}(T)m_{X}(T)\varepsilon _{\mu },
\label{eq:Res}
\end{equation}
with $\varepsilon _{\mu }$ being the polarization vector of the $X(3872)$
state. Then in terms of $m_{X}(T)$ and $f_{X}(T)$, the correlation function
can be written in the form
\begin{equation}
\Pi _{\mu \nu }^{\mathrm{Phys}}(q,T)=\frac{m_{X}^{2}(T)f_{X}^{2}(T)} {%
m_{X}^{2}(T)-q^{2}} \left( -g_{\mu \nu }+\frac{q_{\mu }q_{\nu }}{m_{X}^{2}(T)%
}\right) +\ldots.  \label{eq:CorM}
\end{equation}

The required sum rules can be obtained by using coefficient, $\Pi _{1}^{%
\mathrm{Phys} }(q,T) $, corresponding to the structure $-g_{\mu\nu}$. After
performing the Borel transformation, the physical sides is found as

\begin{eqnarray}
&&\mathcal{B}_{q^{2}}\Pi _{1 }^{\mathrm{Phys}
}(q,T)=m_{X}^{2}(T)f_{X}^{2}(T)e^{-m_{X}^{2}(T)/M^{2}}.  \label{eq:CorBor}
\end{eqnarray}

The correlation function in the QCD side, $\Pi _{\mu \nu
}^{\mathrm{QCD}}(q,T)$, has to be determined employing the
quark-gluon degrees of freedom. To this end, we contract the heavy
and light quark fields and find for the correlation function $\Pi
_{\mu \nu }^{\mathrm{QCD}}(q,T)$ in the diquark-antidiquark
picture the following expression:
\begin{eqnarray}
&&\Pi _{\mu \nu }^{\mathrm{QCD}}(q,T)=-\frac{i}{2}\int d^{4}xe^{iq\cdot
x}\epsilon \tilde{\epsilon}\epsilon ^{\prime }\tilde{\epsilon}^{\prime
}\left\langle \left\{ \mathrm{Tr}\left[ \gamma _{5}\widetilde{S}%
_{q}^{aa^{\prime }}(x)\right. \right. \right.   \notag \\
&&\left. \times \gamma _{5}S_{c}^{bb^{\prime }}(x)\right] \mathrm{Tr}\left[
\gamma _{\mu }\widetilde{S}_{c}^{e^{\prime }e}(-x)\gamma _{\nu
}S_{q}^{d^{\prime }d}(-x)\right]   \notag \\
&&+\mathrm{Tr}\left[ \gamma _{\mu }\widetilde{S}_{c}^{e^{\prime
}e}(-x)\gamma _{5}S_{q}^{d^{\prime }d}(-x)\right] \mathrm{Tr}\left[ \gamma
_{\nu }\widetilde{S}_{q}^{aa^{\prime }}(x)\right.   \notag \\
&&\times \left. \gamma _{5}S_{c}^{bb^{\prime }}(x)\right] +\mathrm{Tr}\left[
\gamma _{5}\widetilde{S}_{q}^{a^{\prime }a}(x)\gamma _{\mu }S_{c}^{b^{\prime
}b}(x)\right]   \notag \\
&&\times \mathrm{Tr}\left[ \gamma _{5}\widetilde{S}_{c}^{e^{\prime
}e}(-x)\gamma _{\nu }S_{q}^{d^{\prime }d}(-x)\right] +\mathrm{Tr}\left[
\gamma _{\nu }\widetilde{S}_{q}^{aa^{\prime }}(x)\right.   \notag \\
&&\left. \left. \times \left. \gamma _{\mu }S_{c}^{bb^{\prime }}(x)\right]
\mathrm{Tr}\left[ \gamma _{5}\widetilde{S}_{c}^{e^{\prime }e}(-x)\gamma
_{5}S_{q}^{d^{\prime }d}(-x)\right] \right\} \right\rangle _{T}.
\label{eq:CorrF2}
\end{eqnarray}%
In Eq.\ (\ref{eq:CorrF2}) we use the notation
\begin{equation*}
\widetilde{S}_{c(q)}^{ij}(x)=CS_{c(q)}^{ijT}(x)C,
\end{equation*}%
with $S_{q}^{ij}(x)$ and $S_{c}^{ij}(x)$ being the light and heavy
quark propagators at the finite temperature, respectively. The
quark propagator in vacuum  was investigated, in detailed,  in
many studies in external spinor  and  gauge fields. In the result
of these investigations, the quark propagator can be written in
terms of quark and gluon condensates \cite{Reinders,Wang:2009ry}.
At finite temperature breakdown of the Lorentz invariance by the
choice of the reference frame and appearance of the residual
$O(3)$ symmetry the new operators arise in operator product
expansions and, therefore, the thermal propagator includes new
terms compared with the vacuum quark propagators
\cite{Mallik:1997pq}.


\begin{eqnarray}
S_{q}^{ij}(x) &=&i\frac{\slashed
x}{2\pi^{2}x^{4}}\delta_{ij}-\frac{
m_{q}}{4\pi^{2}x^{2}}\delta_{ij}  \notag \\
&-&\frac{\langle \bar{q}q\rangle }{12}\delta_{ij}-\frac{x^{2}}{192}%
m_{0}^{2}\langle \bar{q}q\rangle \Big[1-i\frac{m_{q}}{6}\slashed x \Big]%
\delta _{ij}  \notag \\
&+&\frac{i}{3}\Big[\slashed x \Big(\frac{m_{q}}{16}\langle
\bar{q}q\rangle
-\frac{1}{12}\langle u\Theta ^{f}u\rangle \Big)  \notag \\
&+&\frac{1}{3}\Big(u\cdot x\Big)\slashed u \langle u\Theta ^{f}u\rangle %
\Big]\delta _{ij}  \notag \\
&-&\frac{ig_{s}\lambda _{ij}^{A}}{32\pi ^{2}x^{2}}G_{A}^{\mu \nu }
\Big(\slashed x \sigma _{\mu \nu }+\sigma _{\mu \nu }\slashed
x\Big), \label{lightquarkpropagator}
\end{eqnarray}%
where $m_{q}$ denotes the light quark mass, $\langle \bar{q}q\rangle $ is
the temperature-dependent light quark condensate, $G_{A}^{\mu \nu }$ is the
external gluon field, $\Theta _{\mu \nu }^{f}$ is the fermionic part of the
energy momentum tensor and $u_{\mu }$ is the four-velocity of the heat bath.
In Eq.\ (\ref{lightquarkpropagator}) $i,\,j$ are color indexes, $\lambda
_{A}^{ij}$ are the standard Gell-Mann matrices with $A=1,\,2\,\ldots 8$.

For the thermal heavy quark propagator $S_{c}^{ij}(x)$ we employ the
expression
\begin{eqnarray}
&&S_{c}^{ij}(x)=i\int \frac{d^{4}k}{(2\pi )^{4}}e^{-ik\cdot x}\left[ \frac{%
\delta _{ij}\left( {\!\not\!{k}}+m_{c}\right) }{k^{2}-m_{c}^{2}}\right.
\notag \\
&&-\frac{gG_{ij}^{\alpha \beta }}{4}\frac{\sigma _{\alpha \beta }\left( {%
\!\not\!{k}}+m_{c}\right) +\left( {\!\not\!{k}}+m_{c}\right) \sigma _{\alpha
\beta }}{(k^{2}-m_{c}^{2})^{2}}  \notag \\
&&\left. +\frac{g^{2}}{12}G_{\alpha \beta }^{A}G_{A}^{\alpha \beta }\delta
_{ij}m_{c}\frac{k^{2}+m_{c}{\!\not\!{k}}}{(k^{2}-m_{c}^{2})^{4}}+\ldots %
\right],  \label{eq:Qprop}
\end{eqnarray}%
where
\begin{equation*}
G_{ij}^{\alpha \beta }\equiv G_{A}^{\alpha \beta }\lambda_{ij}^{A}/2.
\end{equation*}

The correlation function $\Pi _{\mu \nu }^{\mathrm{QCD}}(q,T)$ can be
decomposed over the Lorentz structures
\begin{eqnarray}
\Pi _{\mu \nu }^{\mathrm{QCD}}(q,T) &=&\Pi _{0}^{\mathrm{QCD}}(q^{2},T)\frac{%
q_{\mu }q_{\nu }}{q^{2}}  \notag \\
&&+\Pi _{1}^{\mathrm{QCD}}(q^{2},T)(-g_{\mu \nu }+\frac{q_{\mu }q_{\nu }}{%
q^{2}}),
\end{eqnarray}%
where $\Pi _{0}^{\mathrm{QCD}}(q^{2},T)$ and $\Pi _{1}^{\mathrm{QCD}%
}(q^{2},T)$ are invariant functions which are related to the
scalar and vector currents, respectively. The QCD sum rule
expressions for the mass and meson-current coupling constant are
derived after fixing the same structures
in both $\Pi _{\mu \nu }^{\mathrm{Phys}}(q,T)$ and $\Pi _{\mu \nu }^{\mathrm{%
QCD}}(q,T)$. As in the physical side of the sum rule, in its QCD side the
structure $-g_{\mu \nu }$ has been taken into account. In the case $%
\overrightarrow{q}=0$, i.e. in the rest frame of the resonance particle, we
can write $\Pi _{1}^{\mathrm{QCD}}(q_{0}^{2},T)$ as the dispersion integral,
\begin{equation}
\Pi _{1}^{\mathrm{QCD}}(q_{0}^{2},T)=\int_{4m_{c}^{2}}^{s_{0}(T)}\frac{\rho
^{\mathrm{QCD}}(s,T)}{s-q_{0}^{2}}ds+...,
\end{equation}%
where $\rho ^{\mathrm{QCD}}(s,T)$ is the corresponding spectral
density. The main question of this section is the calculation of
$\rho ^{\mathrm{QCD}}(s,T)$. In the present work we include into
our sum rules the quark, gluon and mixed condensates up to six
dimensions. For computation of the components of the
spectral densities we use the technical methods presented in Ref.\ \cite%
{Agaev:2016dev}. Results of our calculations are collected in the
appendix. Let us note that we have used the following relation to
express the gluon condensate in terms of the gluonic part of the
energy-momentum tensor $\Theta _{\lambda \sigma }^{g}$ (see for
details Ref. \cite{Mallik:1997pq}):
\begin{eqnarray}
&&\langle Tr^{c}G_{\alpha \beta }G_{\mu \nu }\rangle =\frac{1}{24}(g_{\alpha
\mu }g_{\beta \nu }-g_{\alpha \nu }g_{\beta \mu })\langle G_{\lambda \sigma
}^{a}G^{a\lambda \sigma }\rangle   \notag  \label{TrGG} \\
&&+\frac{1}{6}\Big[g_{\alpha \mu }g_{\beta \nu }-g_{\alpha \nu }g_{\beta \mu
}-2(u_{\alpha }u_{\mu }g_{\beta \nu }-u_{\alpha }u_{\nu }g_{\beta \mu }
\notag \\
&&-u_{\beta }u_{\mu }g_{\alpha \nu }+u_{\beta }u_{\nu }g_{\alpha \mu })\Big]%
\langle u^{\lambda }{\Theta }_{\lambda \sigma }^{g}u^{\sigma }\rangle .
\end{eqnarray}

Applying the Borel transformation on the variable $q_0^{2}$ in the invariant
amplitude $\Pi_{1}^{\mathrm{QCD}}(q^{2},T)$, equating the obtained
expression with the relevant part of $\mathcal{B}_{q^{2}}\Pi _{\mu \nu }^{%
\mathrm{Phys}}(q,T)$, and subtracting the continuum contribution, we finally
obtain the required sum rule. Thus, the mass of the $X(3872)$ state can be
evaluated from the sum rule
\begin{equation}
m_{X}^{2}(T)=\frac{\int_{4m_{c}^{2}}^{s_{0}(T)}dss\rho ^{\mathrm{QCD}%
}(s,T)e^{-s/M^{2}}}{\int_{4m_{c}^{2}}^{s_{0}(T)}ds\rho^{\mathrm{QCD}}
(s,T)e^{-s/M^{2}}},  \label{eq:srmass}
\end{equation}%
whereas to extract the numerical value of the meson-current coupling
constant $f_{X}(T)$ we employ the formula
\begin{equation}
f_{X}^{2}(T)e^{-m_{X}^{2}(T)/M^{2}}=\frac{1}{m_{X}^{2}(T)}
\int_{4m_{c}^{2}}^{s_{0}(T)}ds\rho ^{\mathrm{QCD}}(s,T)e^{-s/M^{2}}.
\label{eq:srcoupling}
\end{equation}

\section{Numerical Analysis}

\label{sec:Num}

The QCD sum rules for the mass and coupling constant of the X(3872) state at
finite temperature contain as parameters various quark, gluon and mixed
vacuum condensates. Their values are collected in Table \ref{tab:Param}.
\begin{table}[tbp]
\begin{tabular}{|c|c|}
\hline\hline
Parameters & Values \\ \hline\hline
$m_{c}$ & $1.28 \pm 0.03 ~\mathrm{GeV} $ \cite{Patrignani} \\
$\langle \bar{q}q \rangle $ & $(-0.24\pm 0.01)^3 ~\mathrm{GeV}^3$ \cite%
{Shifman,Reinders} \\
$\langle\frac{\alpha_sG^2}{\pi}\rangle $ & $(0.012\pm0.004)~\mathrm{GeV}^4$
\cite{Shifman,Reinders} \\
$m_0^2 $ & $(0.8\pm0.1) ~\mathrm{GeV}^2 $ \cite{Shifman,Reinders}  \\
\hline\hline
\end{tabular}%
\caption{Input parameters.}
\label{tab:Param}
\end{table}

Beside these parameters, we need the temperature dependent quark
and gluon condensates, as well as the temperature dependent energy
density. In the case of quark condensate, we use the fit function
obtained in Refs. \cite{Azizi:2016ddw,Ayala}, which coincides with
the lattice QCD results \cite{Bazavov,Cheng1}
\begin{equation*}
\langle \bar{q}q\rangle =\frac{\langle 0|\bar{q}q|0\rangle }{%
1+e^{18.10042(1.84692[\frac{1}{\mathrm{GeV}^{2}}]T^{2}+4.99216[\frac{1}{%
\mathrm{GeV}}]T-1)}},
\end{equation*}%
and is valid up to a critical temperature $T_{c}=190~\mathrm{MeV}$ and $%
\langle 0|\bar{q}q|0\rangle $ is the vacuum condensate of the light quarks.

For the gluonic and fermionic parts of the energy density we use
the  parametrization obtained in Ref. \cite{Azizi:2016ddw} from
the lattice QCD graphics presented in Ref.  \cite{Cheng:2007jq}:
\begin{eqnarray}
\langle \Theta _{00}^{g}\rangle =\dfrac{1}{2}\langle \Theta _{00}\rangle
&=&T^{4}e^{(113.867[\frac{1}{GeV^{2}}]T^{2}-12.190[\frac{1}{GeV}]T)}  \notag
\label{tetamumu} \\
&-&10.141[\frac{1}{GeV}]T^{5}.
\end{eqnarray}

Finally, we use the parametrization for the temperature-dependent
gluon condensate obtained via QCD sum rules  and lattice QCD data
\cite{Azizi:2016ddw,Ayala2}:
\begin{eqnarray}
\langle G^{2}\rangle  &=&\langle 0|G^{2}|0\rangle \Bigg[1-1.65\Big(\frac{T}{%
T_{c}}\Big)^{8.735}  \notag  \label{G2TLattice} \\
&+&0.04967\Big(\frac{T}{T_{c}}\Big)^{0.7211}\Bigg],
\end{eqnarray}%
where $\langle 0|G^{2}|0\rangle $ is the gluon condensate in the vacuum.

The temperature-dependent continuum threshold for the $X(3872)$
state is one of the auxiliary parameters that should also be
determined. We used the continuum threshold in terms of
temperature \cite{Dominguez:2009mk,Veliev:2010gb}
\begin{eqnarray}  \label{G2TLattice}
s_0(T)&=&s_0\left[1-\left(\frac{T}{T_c}\right)^8\right] +4m_c^2\left(\frac{T%
}{T_c}\right)^8,
\end{eqnarray}
where $s_0$ is the continuum threshold at $T=0$. This parameter is
not arbitrary and depends on the energy of the first excited state
with the same quantum numbers as the chosen interpolating currents
for the $X(3872)$ state. For $s_0$, we take the interval
\begin{equation}
16.5~\mathrm{GeV}^2\leq s_0\leq 17.0~\mathrm{GeV}^2,  \label{eq:19}
\end{equation}
in which the physical quantities show relatively weak dependence on it.

According to the philosophy of the used method, the physical
quantities should be practically independent the auxiliary
parameters $M^2$. Also, in order to fix the working window for the
Borel parameter $M^2$, we require the convergence of the OPE, as
well as the suppression of the contributions arising from the
higher resonances and continuum, in other words exceeding of the
pole contribution over the ones coming from the higher dimensional
condensates. As a result, for the mass and coupling constant
calculations we find the range of $M^2$
\begin{equation}
3.5~\mathrm{GeV}^2\leq M^2\leq 5.5~\mathrm{GeV}^2,  \label{eq:19â}
\end{equation}
as reliable for our purposes. It is worth noting that in this
interval the dependence of the mass and meson-current coupling
constant on $M^2$ is stable, and we may expect that the sum rules
give the correct results. In order to demonstrate independence of
physical quantities from $M^2$ and $s_0$, we plot the mass and
meson-current coupling constant versus $M^2$ at different fixed
values of the continuum threshold $s_0$ at $T = 0$ and vice versa
in Figs.\ \ref{fig1} and \ref{fig2}. From these figures, we see
that these quantities depend on both $M^2$ and $s_0$ very weakly
in their working intervals.

\begin{widetext}

\begin{figure}[ht]
\begin{center}
\subfigure[]{\includegraphics[width=8cm]{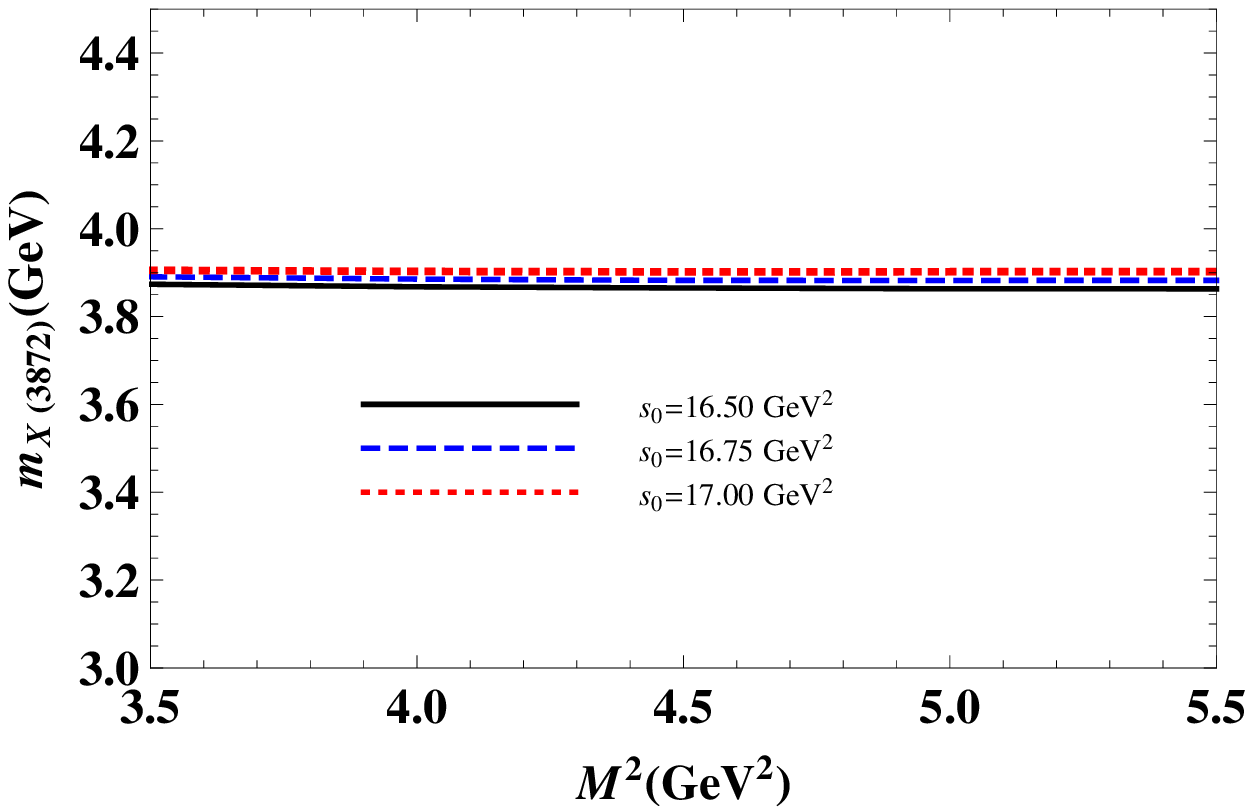}}
\subfigure[]{\includegraphics[width=8cm]{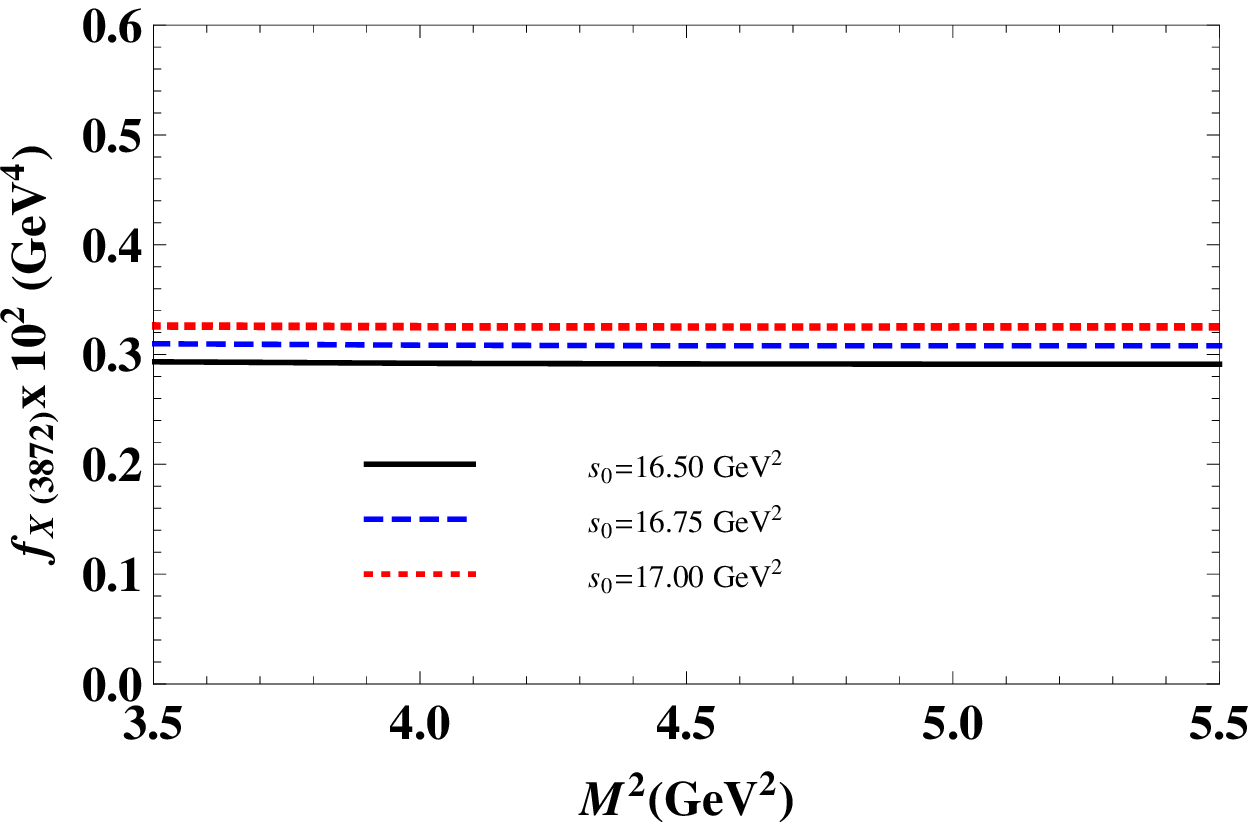}}
\end{center}
\caption{(a) The mass of the $X(3872)$ state as a function of
$M^2$ for different fixed values of $s_0$ at $T=0$. (b)  The same
as (a) but for the coupling parameter $f_{X}$.} \label{fig1}
\end{figure}
\begin{figure}[ht]
\begin{center}
\subfigure[]{\includegraphics[width=8cm]{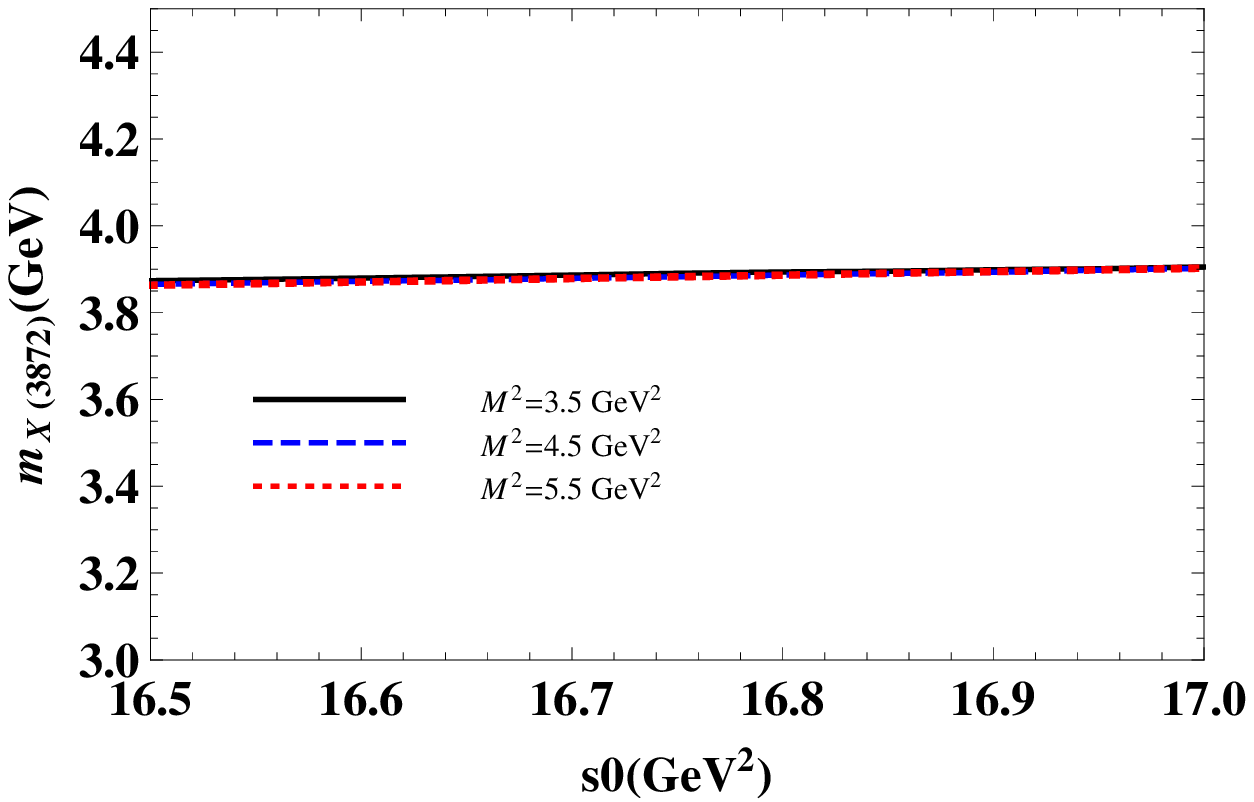}}
\subfigure[]{\includegraphics[width=8cm]{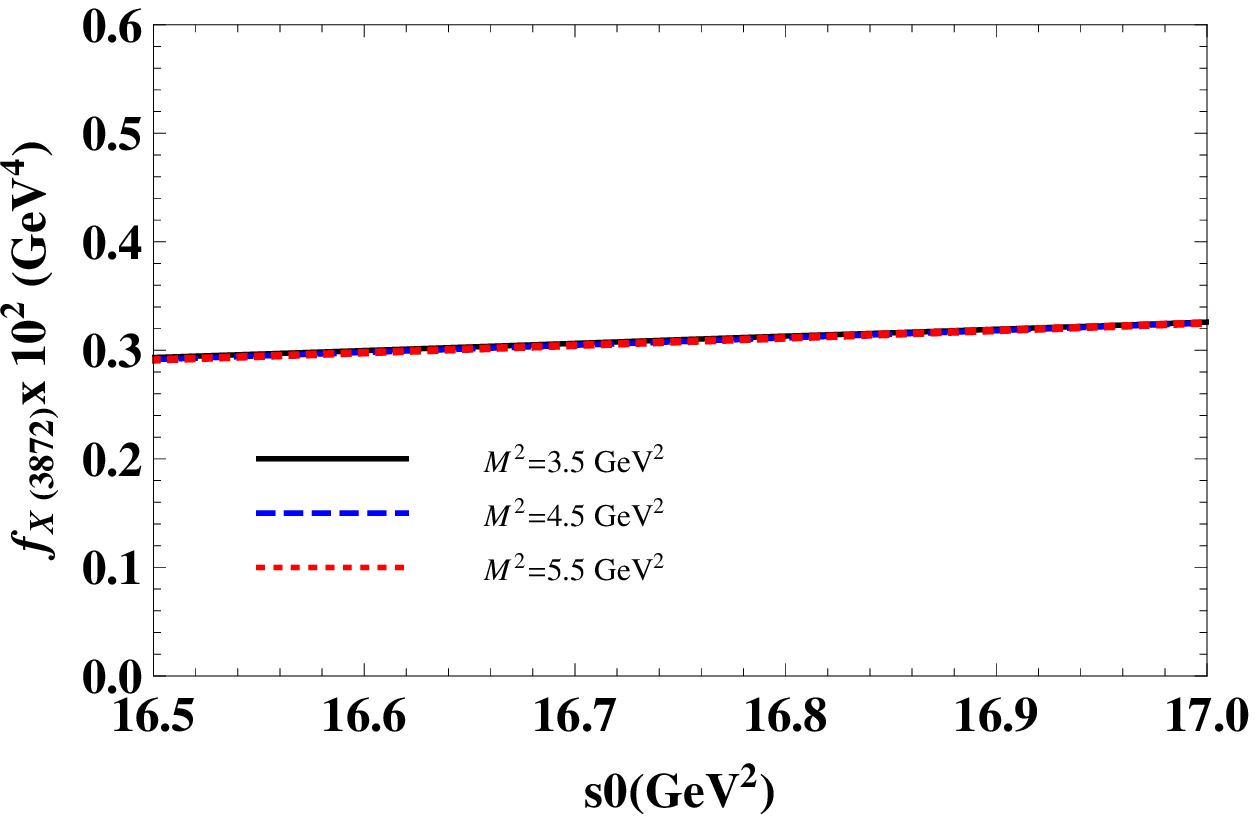}}
\end{center}
\caption{(a) The mass of the $X(3872)$ state as a function of
$s_0$ for different fixed values of $M^2$ at $T=0$. (b)  The same
as (a) but for the coupling parameter $f_{X}$.} \label{fig2}
\end{figure}

\end{widetext}

\begin{widetext}

\begin{figure}[ht]
\begin{center}
\subfigure[]{\includegraphics[width=8cm]{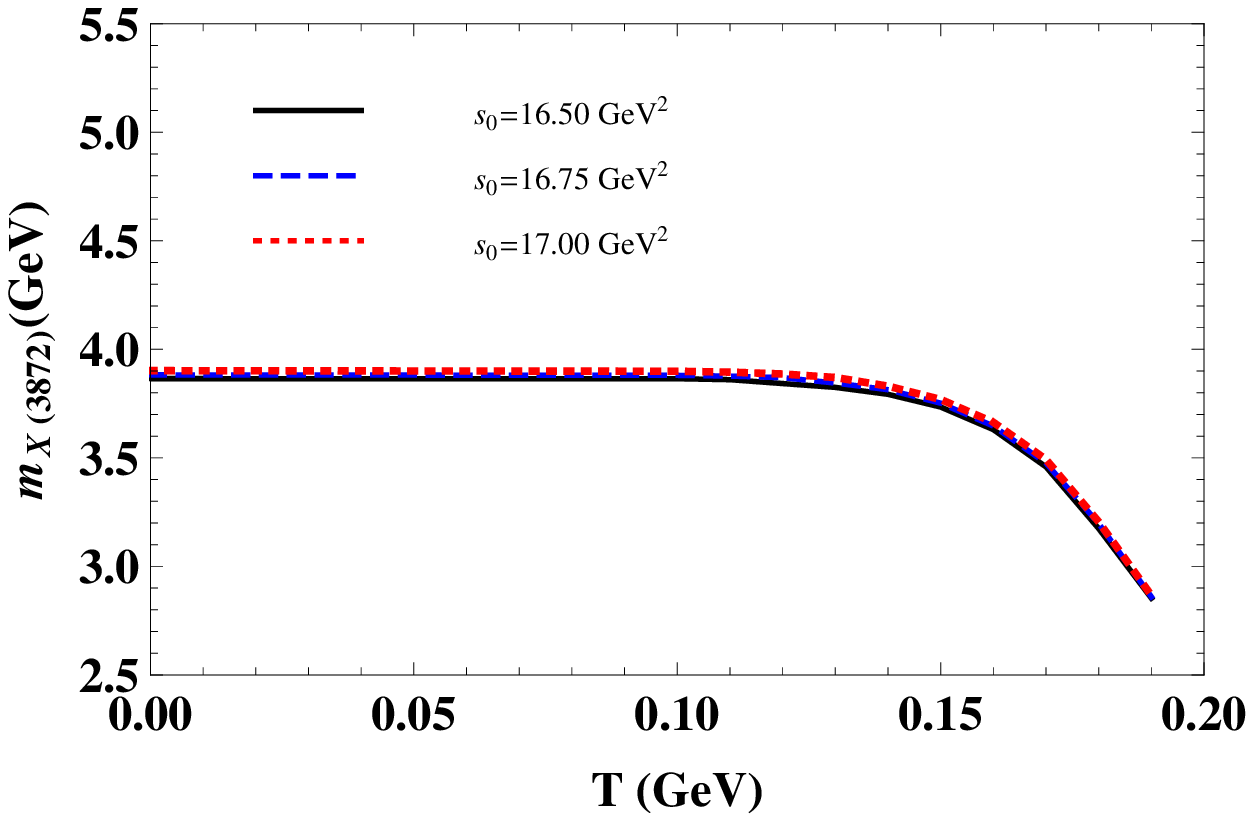}}
\subfigure[]{\includegraphics[width=8cm]{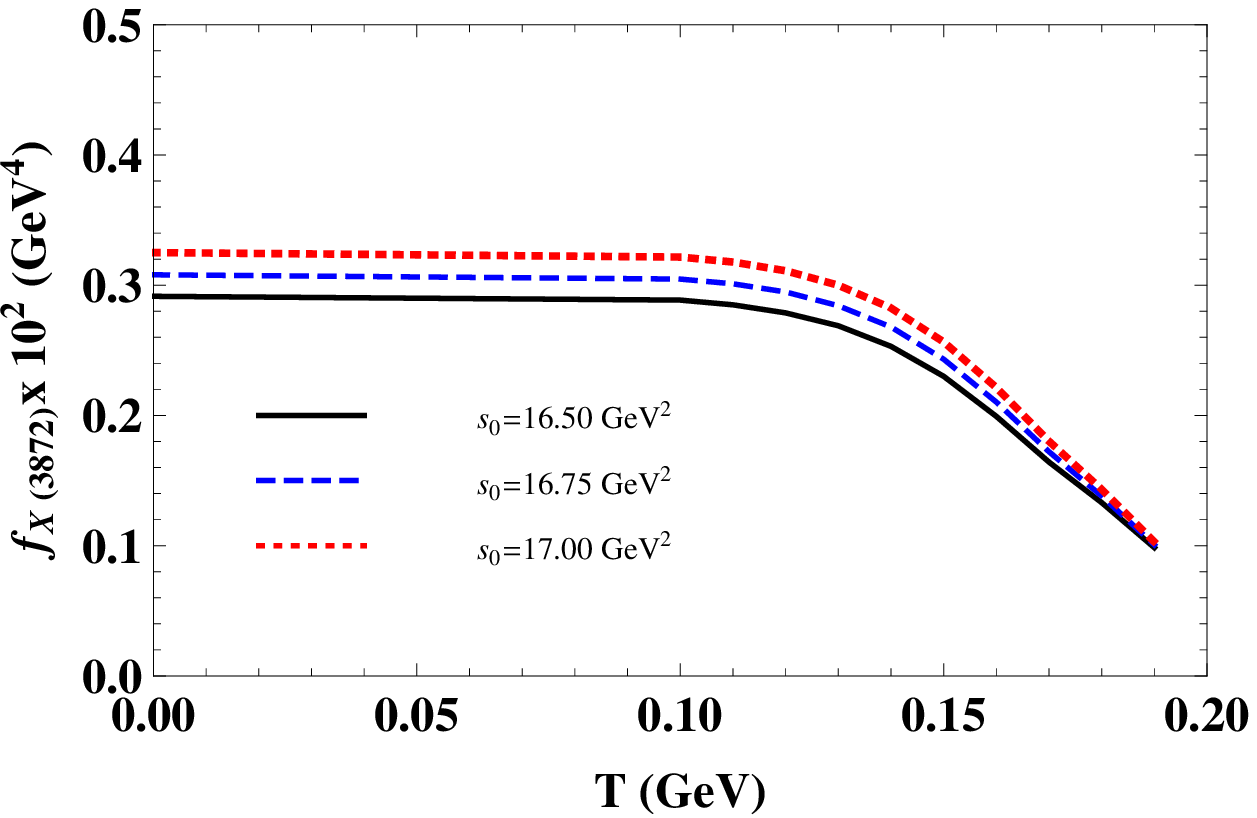}}
\end{center}
\caption{(a) The mass of the $X(3872)$ state as a function of
temperature  for different values of $s_0$ and at fixed value of
$M^2=4.5~\mathrm{GeV}^2$. (b) The same as (a) but for the coupling
parameter $f_{X}$.} \label{fig3}
\end{figure}

\end{widetext}

The final task is to investigate the variations of the mass and
coupling parameter of the $X(3872)$ state with respect to
temperature. For this purpose, we plot these quantities as a
function of temperature in Fig.\ \ref{fig3}. This figure indicates
that the mass and coupling constant of the $X(3872)$ state remain
approximately unchanged up to $T \cong 0.11\ \mathrm{GeV}$ ,
however, after this point, they start to diminish rapidly by
increasing the temperature. Near the critic or deconfinement
temperature, the coupling constant reaches approximately $34\%$ of
its value in vacuum, while the mass decreased by $26\%$. From this
figure we deduce the result on the meson-current coupling constant
and mass in vacuum as presented in Table II. In the case of the
mass, our result within the uncertainties is in good agreement
with those of \cite{Patrignani,Sundu:2016oda}. On the other hand,
the result of the meson-current coupling is smaller than the
result in the literature \cite{Sundu:2016oda}. It can be checked
in the future experiments.

\begin{table}[tbp]
\begin{tabular}{|c|c|c|}
\hline\hline
& $m_{X(3872)}(\mathrm{MeV})$ & $f_{X(3872)}\times 10^{2} (\mathrm{GeV^{4}})$
\\ \hline\hline
Present Work & $3885\pm85$ & $0.31\pm0.12$ \\ \hline
Experiment \cite{Patrignani} & $3871.69\pm0.17$ & - \\ \hline
\cite{Sundu:2016oda} & $3873\pm127 $ & $0.56\pm0.19 $ \\ \hline\hline
\end{tabular}%
\caption{Values of mass and coupling constant of the $X(3872) $ state in the
vacuum.}
\label{tab:Results1}
\end{table}

\section{Conclusions}

In the present work we have investigated the diquark-antidiquark
$X(3872)$ meson by calculating its spectroscopic parameters in the
framework of the thermal QCD sum rules method. The analysis of the
obtained thermal sum rule allows us to study contributions of a
medium to the mass and coupling constant of the $X(3872)$
resonance. Our numerical calculations demonstrate that the mass
and meson-current coupling constant are insensitive to the
variation of the temperature up to $T=110~MeV$ , however after
this point; they start to fall by increasing of the temperature.
At deconfinement temperature, the meson-current coupling constants
attain roughly to $34\%$ of its vacuum value. But decreasing of
the mass and current coupling with the temperature does not mean a
stability of the particle under consideration. To make a
conclusion about the stability of the particle one has to
calculate its decay width. Indeed, apart from the mass and
coupling constant, the  decay width of the particle depends also
on other temperature-dependent parameters. In
Ref.\cite{Azizi:2010zza} by explicit calculations of the decay
width, it was shown that, despite decreasing of the mass and
coupling constant of pseudoscalar particles their decay widths
increase with the temperature revealing their unstable nature.
Therefore, decreasing the mass and coupling of the particle with
temperature does not automatically lead to decreasing its width.
In the future our aim is to investigate the temperature dependent
decay width of the $X(3872)$ state.

The considerable decrease in the values of mass and coupling
parameter  can be considered as a sign of the quark-gluon plasma
 phase  transition. Also, the obtained behavior  in
terms of temperature can be used in the analysis of the heavy-ion
collision experiments. Our predictions for the spectroscopic
properties of the $X(3872)$ state can be checked in future
experiments.

\section*{Acknowledgment}

\label{sec:e}

The authors thank Kocaeli University for the partial financial support
through the grant BAP 2017/018.

\appendix*

\section{ The two-point thermal spectral density $\protect\rho^{\mathrm{QCD}%
}(s, T)$}

\renewcommand{\theequation}{\Alph{section}.\arabic{equation}} \label{sec:App}

In this appendix we have collected the results of our calculations of the
spectral density
\begin{eqnarray}
\rho^{\mathrm{QCD}}(s,T)=\rho^{\mathrm{pert.}}(s)+\sum_{k=3}^{6}\rho
_{k}(s,T),  \label{eq:A1}
\end{eqnarray}%
necessary for the evaluation of the mass and coupling constant of
the temperature-dependent $X(3872)$ state from the QCD sum rules.
$\rho_{k}(s,T)$
denote the nonperturbative contributions to $\rho^{QCD}(s,T)$ and $%
g_s=4\pi\alpha_s$. The explicit expressions for $\rho^{\mathrm{pert.}}(s)$
and $\rho_{k}(s,T)$ are presented below as the integrals over the Feynman
parameters $z$ and $w$:
\begin{eqnarray}
\rho^{\mathrm{pert.}}(s)&=&\frac{1}{3072\pi^6}\int_{0}^{1}dz\int_{0}^{1-z}dw%
\frac{wz}{ht^8}  \notag \\
&\times& \left[swzh-m_c^2t(w+z)\right]^2  \notag \\
&\times&[35h^2w^2z^2s^2-26htwz(w+z)sm_c^2  \notag \\
&+&3t^2(w+z)^2m_c^2] \theta[L],
\end{eqnarray}
\begin{eqnarray}
\rho^{3}(s,T)&=&\frac{\langle \bar{u}u\rangle m_c}{64\pi^4}%
\int_{0}^{1}dz\int_{0}^{1-z}dw\frac{\left[t(w+z)m_c^2-hwzs\right] }{t^5}
\notag \\
&\times&(w+z)\left[7swzh-3m_c^2t(w+z)\right] \theta[L]  \notag \\
&-&\frac{\langle \bar{u}\Theta^{f}u\rangle }{12\pi^4}\int_{0}^{1}dz%
\int_{0}^{1-z}dw\frac{hwz }{t^6}\left[15h^2s^2w^2z^2\right.  \notag \\
&-&\left. 13hm_c^2stwz(w+z)+m_c^4t^2(w+z)^2\right] \theta[L],  \notag \\
&&{}
\end{eqnarray}
\begin{eqnarray}
\rho^{4}(s,T)&=&\frac{1}{36864\pi^{4}}\langle\alpha_{s}\frac{G^2}{\pi}%
\rangle \int_{0}^{1}dz\int_{0}^{1-z}dw\frac{w}{h^2t^6}  \notag \\
&\times&\Big [480h^4s^2w^2z^4-hm_c^2stwz\Big(60w^2(w-1)^3  \notag \\
&+& w(w-1)z(120+w(353w-345)) +z^2(879w  \notag \\
&+& w^2(907w-15989)-60)+z^3(w-1)(1577w  \notag \\
&-& 594)+4z^4(321w-220)+346z^5\Big) +m_c^4t^2  \notag \\
&\times& \Big(60w^3(w-1)^3+3w^2(w-1)z(60-155w  \notag \\
&+& 111w^2)+4wz^2(240w+w^2(179w-362)-45)  \notag \\
&+& z^3(705w+w(868w-1465)-60)+z^4(637w^2  \notag \\
&-& 827w+210)+2z^5(127w-96)+42z^6\Big)\Big] \theta[L]  \notag \\
&+& \frac{\alpha_{s}\langle u \Theta^f u\rangle}{9216\pi^{5}}
\int_{0}^{1}dz\int_{0}^{1-z}dw\frac{w}{h^2t^6}\Big[240h^4s^2w^2z^4  \notag \\
&-& 3m_c^2t^2\Big(20w^3(w-1)+15w^2z (w-1)^2(7w-4)  \notag \\
&+& 4wz^2(w-1)(64w^2-74w +15)+z^3(w-1)  \notag \\
&\times& (20-257w+384w^2)+z^4(86-411w+361w^2)  \notag \\
&+& 2z^5(97w-56)+46z^6\Big)+m_c^2htwz \Big(-4sz  \notag \\
&\times& z(-15w^2(w-1)^2+wz(w-1)(109w-129)  \notag \\
&+&z^2(w-1)(379w-144)+4z^3(103w-72)  \notag \\
&+& 144z^2)+15s(4w^2(w-1)^3 +wz(w-1)^2  \notag \\
&\times& (17w-8)+z^2(w-1)(63w^2-63w+4)  \notag \\
&+& z^3(w-1)(129w-46)+4z^4(29w-20)  \notag \\
&+& 38z^5)\Big)\Big]\theta[L]
\end{eqnarray}
\begin{eqnarray}
\rho^{5}(s,T)&=&\frac{m_cm_0^2\langle \overline{u}u\rangle}{128\pi^4}%
\int_{0}^{1}dz\int_{0}^{1-z}dw\frac{wzh(w+z) }{t^5}  \notag \\
&\times& (5shwz-3m_c^2t(w+z))\theta[L],
\end{eqnarray}
\begin{eqnarray}
&&\rho^{6}(s,T)=\frac{1}{108\pi^2}\int_{0}^{1}dz\int_{0}^{1-z}dw\Big[%
9m_c^2\langle \overline{u}u\rangle^2  \notag \\
&-& 3m_c\langle \overline{u}u\rangle \langle u \Theta^f u\rangle+20z \langle
u \Theta^f u\rangle^2 \Big]\theta[L].
\end{eqnarray}
In the expressions above we have used the notations:
\begin{eqnarray}
L&=&\frac{[m_c^2(w^3+w^2(2z-1)+(z^2+2wz)(z-1))-swzh]}{t^2}  \notag \\
&\times& (w-1),  \notag \\
\Phi&=&\frac{m_c^2[w^3+z(z-1)(2w+z)+w^2(2z-1)]}{wzh},  \notag \\
t&=&w^2+(z-1)(w+z),\ h=w+z-1.  \label{eq:A4}
\end{eqnarray}


\begin{thebibliography}{999}

\bibitem{Choi:2003ue}  S.~K.~Choi \textit{et al.} [Belle Collaboration],
Phys.\ Rev.\ Lett.\ \textbf{91}, 262001 (2003).


\bibitem{Acosta:2003zx}  D.~Acosta \textit{et al.} [CDF Collaboration],
Phys.\ Rev.\ Lett.\ \textbf{93}, 072001 (2004).


\bibitem{Abazov:2004kp}  V.~M.~Abazov \textit{et al.} [D0 Collaboration],
Phys.\ Rev.\ Lett.\ \textbf{93}, 162002 (2004).


\bibitem{Adachi:2011ji}  I.~Adachi \textit{et al.} [Belle Collaboration],
Phys.\ Rev.\ Lett.\ \textbf{108}, 032001 (2012).


\bibitem{Abe:2007tk}  K.~F.~Chen \textit{et al.} [Belle Collaboration],
Phys.\ Rev.\ Lett.\ \textbf{100}, 112001 (2008).


\bibitem{Aaij:2015tga}  R.~Aaij \textit{et al.} [LHCb Collaboration],
Phys.\ Rev.\ Lett.\ \textbf{115}, 072001 (2015).


\bibitem{Swanson:2006st}  E.~S.~Swanson,
Phys.\ Rept.\ \textbf{429}, 243 (2006).

\bibitem{Richard}  J.M. Richard, Nucl. Phys (Proc. Suppl.) B164 (2007) 131.


\bibitem{Zhu:2007wz}  S.~L.~Zhu,  
Int.\ J.\ Mod.\ Phys.\ E \textbf{17}, 283 (2008).


\bibitem{Godfrey:2008nc}  S.~Godfrey and S.~L.~Olsen,
Ann.\ Rev.\ Nucl.\ Part.\ Sci.\ \textbf{58}, 51 (2008).


\bibitem{Nielsen:2009uh}  M.~Nielsen, F.~S.~Navarra and S.~H.~Lee,
Phys.\ Rept.\ \textbf{497}, 41 (2010).  


\bibitem{Chen:2010ze}  W.~Chen and S.~L.~Zhu,
Phys.\ Rev.\ D \textbf{83}, 034010 (2011).


\bibitem{Brambilla:2010cs}  N.~Brambilla \textit{et al.},
Eur.\ Phys.\ J.\ C \textbf{71}, 1534 (2011).


\bibitem{Albuquerque:2013ija}  R.~M.~Albuquerque,
PhD Thesis, Univ. S\~ao Paulo (Brazil) and Univ. Montpellier 2  (France),
arXiv:1306.4671 [hep-ph].  


\bibitem{Liu:2013waa}  X.~Liu,  
Chin.\ Sci.\ Bull.\ \textbf{59}, 3815 (2014).


\bibitem{Bodwin:2013nua}  G.~T.~Bodwin, E.~Braaten, E.~Eichten, S.~L.~Olsen,
T.~K.~Pedlar and J.~Russ,
FERMILAB-CONF-13-665-T,  arXiv:1307.7425 [hep-ph].

\bibitem{Nielsen}  M. Nielsen, Nucl. Part. Phys. Proc. 258-259 (2015) 139.

\bibitem{Braaten}  E. Braaten, EPJ Web Conf 113 (2016) 01015.


\bibitem{Esposito:2014rxa}  A.~Esposito, A.~L.~Guerrieri, F.~Piccinini,
A.~Pilloni and A.~D.~Polosa,  
Int.\ J.\ Mod.\ Phys.\ A \textbf{30}, 1530002 (2015)

\bibitem{Chen}  H.-X. Chen et al., Phys. Rep. 631 (2016) 1.

\bibitem{Briceno}  R.A. Briceno et al.,
Chin. Phys. C40 n.4, 042001 (2016).


\bibitem{Wang:2016mmg}  Z.~G.~Wang,
Eur.\ Phys.\ J.\ C \textbf{76}, 387 (2016).


\bibitem{Ali:2016gli}  A.~Ali,
DESY 16-090, arXiv:1605.05954 [hep-ph].  


\bibitem{Wang:2016gxp}  Z.~G.~Wang,
Eur.\ Phys.\ J.\ C \textbf{77}, 78 (2017).


\bibitem{Wang:2016tzr}  Z.~G.~Wang,
Eur.\ Phys.\ J.\ C \textbf{76}, 657 (2016).


\bibitem{Aubert:2004ns}  B.~Aubert \textit{et al.} [BaBar Collaboration],
Phys.\ Rev.\ D \textbf{71}, 071103 (2005).  


\bibitem{Aubert:2005rm}  B.~Aubert \textit{et al.} [BaBar Collaboration],
Phys.\ Rev.\ Lett.\ \textbf{95}, 142001 (2005).


\bibitem{Aaij:2014ala}  R.~Aaij \textit{et al.} [LHCb Collaboration],
Nucl.\ Phys.\ B \textbf{886}, 665 (2014).

\bibitem{Shifman}  M. A. Shifman, A. I. Vainshtein and V. I. Zakharov,
Nucl. Phys. B 147, 385 (1979).

\bibitem{Reinders} L. J. Reinders, H. Rubinstein and S. Yazaki,
Phys. Rept. 127, 1(1985).


\bibitem{Colangelo:2000dp}  P.~Colangelo and A.~Khodjamirian,
In *Shifman, M. (ed.): At the frontier of particle physics, vol. 3*
1495-1576  


\bibitem{Narison:2002pw}  S.~Narison,
Camb. Monogr. Part. Phys. Nucl. Phys. Cosmol. 17:1, 2002.  


\bibitem{Bochkarev} A.I. Bochkarev, M.E. Shaposhnikov, Nucl. Phys. B \textbf{%
268}, 220 (1986).

\bibitem{Hatsuda} T. Hatsuda, Y. Koike, S. H. Lee, Nucl. Phys. B \textbf{394}%
, 221 (1993).

\bibitem{Morita} K. Morita and S. H. Lee, Phys. Rev. Lett. \textbf{100},
022301 (2008).

\bibitem{Gubler} P. Gubler, K. Morita and M. Oka, Phys. Rev. Lett. \textbf{%
107}, 092003 (2011).


\bibitem{Azizi:2015oxa}  K.~Azizi and G.~Kaya,
J.\ Phys.\ G \textbf{43}, no. 5, 055002 (2016).


\bibitem{Azizi:2016ddw}  K.~Azizi and G.~Bozk{\i}r,
Eur.\ Phys.\ J.\ C \textbf{76}, no. 10, 521 (2016).

\bibitem{Azizi:2015ona}  K.~Azizi and G.~Kaya,
Eur.\ Phys.\ J.\ Plus \textbf{130}, no. 8, 172 (2015)


\bibitem{Azizi:2015qva}  K.~Azizi and N.~Kat{\i}rc{\i},
Eur.\ Phys.\ J.\ Plus \textbf{131}, no. 5, 163 (2016)


\bibitem{Azizi:2010zza}  K.~Azizi and N.~Er,
Phys.\ Rev.\ D \textbf{81}, 096001 (2010).  


\bibitem{Veliev:2010gb}  E.~V.~Veliev, H.~Sundu, K.~Azizi and M.~Bayar,
Phys.\ Rev.\ D \textbf{82}, 056012 (2010).


\bibitem{Azizi:2014maa}  K.~Azizi, A.~T\"{u}rkan, E.~Veli Veliev and
H.~Sundu,
Adv.\ High Energy Phys.\ \textbf{2015}, 794243 (2015)


\bibitem{Veliev:2014tca}  E.~V.~Veliev, K.~Azizi, H.~Sundu and G.~Kaya,
Rom.\ J.\ Phys.\ \textbf{59-63}, no. 1-2, 140 (2014).


\bibitem{Yazici:2015tqa}  E.~Yazici, H.~Sundu and E.~V.~Veliev,
Eur.\ Phys.\ J.\ C \textbf{76}, no. 2, 89 (2016)


\bibitem{Veliev:2008zi}  E.~V.~Veliev and T.~M.~Aliev,
J.\ Phys.\ G \textbf{35}, 125002 (2008)



\bibitem{Agaev:2017foq}  S.~S.~Agaev, K.~Azizi and H.~Sundu,
Phys.\ Rev.\ D \textbf{95}, no. 11, 114003 (2017)


\bibitem{Agaev:2017uky}  S.~S.~Agaev, K.~Azizi and H.~Sundu,
Eur.\ Phys.\ J.\ C \textbf{77}, no. 5, 321 (2017)


\bibitem{Agaev:2016dsg}  S.~S.~Agaev, K.~Azizi and H.~Sundu,
Phys.\ Rev.\ D \textbf{95}, no. 3, 034008 (2017)


\bibitem{Agaev:2016ifn}  S.~S.~Agaev, K.~Azizi, B.~Barsbay and H.~Sundu,
Eur.\ Phys.\ J.\ A \textbf{53}, no. 1, 11 (2017)


\bibitem{Agaev:2016srl}  S.~S.~Agaev, K.~Azizi and H.~Sundu,
Phys.\ Rev.\ D \textbf{93}, no. 11, 114036 (2016)


\bibitem{Sundu:2016oda}
  H.~Sundu,
  SDU J.\ Nat.\ Appl.\ Sci.\  {\bf 20}, no. 3, 448 (2016).



\bibitem{Abreu:2017nuc}
  L.~M.~Abreu, K.~P.~Khemchandani, A.~Martínez Torres, F.~S.~Navarra, M.~Nielsen and A.~L.~Vasconcellos,
  Phys.\ Rev.\ D {\bf 95}, no. 9, 096002 (2017)



\bibitem{Cho:2017dcy}  S.~Cho \textit{et al.} [ExHIC Collaboration],
Prog.\ Part.\ Nucl.\ Phys.\ \textbf{95}, 279 (2017)


\bibitem{Abreu:2016kbd}  L.~M.~Abreu, K.~P.~Khemchandani, A.~M.~Torres,
F.~S.~Navarra and M.~Nielsen,
J.\ Phys.\ Conf.\ Ser.\ \textbf{736}, no. 1, 012026 (2016).


\bibitem{Abreu:2016qci}  L.~M.~Abreu, K.~P.~Khemchandani, A.~Martinez
Torres, F.~S.~Navarra and M.~Nielsen,
Phys.\ Lett.\ B \textbf{761}, 303 (2016)


\bibitem{Wang:2017jtz}  Z.~G.~Wang,
Eur.\ Phys.\ J.\ C \textbf{77}, no. 7, 432 (2017)


\bibitem{Wang:2016dcb}  Z.~G.~Wang,
Eur.\ Phys.\ J.\ C \textbf{77}, no. 3, 174 (2017)


\bibitem{Wang:2016wwe}  Z.~G.~Wang,
Chin.\ Phys.\ C \textbf{41}, 083103 (2017)



\bibitem{Matheus:2006xi}  R.~D.~Matheus, S.~Narison, M.~Nielsen and
J.~M.~Richard,  
Phys.\ Rev.\ D \textbf{75}, 014005 (2007).


\bibitem{Wang:2009ry}  Z.~G.~Wang, Z.~C.~Liu and X.~H.~Zhang,
Eur.\ Phys.\ J.\ C \textbf{64}, 373 (2009)









\bibitem{Mallik:1997pq}  S.~Mallik,
Phys.\ Lett.\ B \textbf{416}, 373 (1998)


\bibitem{Agaev:2016dev}  S.~S.~Agaev, K.~Azizi and H.~Sundu,
Phys.\ Rev.\ D \textbf{93}, 074002 (2016).

\bibitem{Patrignani}  C. Patrignani et al. (Particle Data Group),
Chin. Phys. C, 40, 100001 (2016) and 2017 update.

\bibitem{Ayala} A. Ayala, A. Bashir, C. A. Dominguez, E. Gutierrez, M.
Loewe, A. Raya, 
Phys. Rev. D \textbf{84}, 056004 (2011).

\bibitem{Bazavov} A. Bazavov et al.,
Phys. Rev. D \textbf{80}, 014504 (2009).

\bibitem{Cheng1} M. Cheng et al.,
Phys. Rev. D \textbf{81}, 054504 (2010).


\bibitem{Cheng:2007jq}  M.~Cheng \textit{et al.},
Phys.\ Rev.\ D \textbf{77}, 014511 (2008)  doi:10.1103/PhysRevD.77.014511
[arXiv:0710.0354 [hep-lat]].

\bibitem{Ayala2} A. Ayala, C. A. Dominguez, M. Loewe, Y. Zhang,
Phys. Rev. D \textbf{86}, 114036 (2012).


\bibitem{Dominguez:2009mk}  C.~A.~Dominguez, M.~Loewe, J.~C.~Rojas and
Y.~Zhang,
Phys.\ Rev.\ D \textbf{81}, 014007 (2010).


\end{thebibliography}
\end{document}